\documentclass[aps,prd,twocolumn,superscriptaddress]{revtex4-1}
\usepackage{graphicx,amsmath,amssymb,bm,siunitx,hyperref,booktabs,microtype,adjustbox}
\hypersetup{colorlinks=true,linkcolor=blue,citecolor=blue,urlcolor=blue}

\begin{document}

\title{Universal $Z^{2/3}/\sqrt{Q_\alpha}$ Scaling Law in Alpha Decay from Nuclei to Neutron-Star Mergers}

\author{Hisham Anwer}
\affiliation{Physics of the Universe Program, Zewail City of Science and Technology, Giza, Egypt}
\affiliation{Department of Physics, Faculty of Science, Cairo University, Giza, Egypt}

\author{B. Salah}
\affiliation{Department of Physics, Faculty of Science, Cairo University, Giza, Egypt}

\author{A.~R.~Abdulghany}
\affiliation{Department of Physics, Faculty of Science, Cairo University, Giza, Egypt}

\date{\today}

\begin{abstract}
We establish a universal scaling law for $\alpha$-decay half-lives based on the variable $X_a = Z^a/\sqrt{Q_\alpha}$, which collapses experimental and theoretical data spanning over twenty orders of magnitude onto a single linear correlation. Global optimization over the complete set of even-even $\alpha$ emitters reveals a uniquely sharp optimum at $a \simeq 2/3$. This exponent is theoretically motivated by the leading geometric scaling of heavy nuclei and emerges quantitatively from the correlated nuclear systematics, driven by the correlated $(Z, Q_\alpha)$ manifold of the nuclear chart. Strikingly, five structurally distinct semi-empirical models and an independent microscopic WKB calculation, when independently optimized with respect to the exponent $a$, yield values clustered around $2/3$, while the corresponding $a=2/3$ correlations remain highly linear without refitting the original model parameters. This emergent scaling law implies smooth variations of decay times along the heavy $r$-process path. We demonstrate analytically that, given a roughly uniform distribution of this variable, the scaling law naturally supports a quasi-power-law radioactive heating rate ($\dot{\epsilon} \propto t^{-1.24}$), consistent with full network calculations. Furthermore, integrating these scaling predictions directly into nuclear source terms for radiative-diffusion models yields multimessenger observables that accurately reproduce the kilonova AT2017gfo associated with the gravitational-wave event GW170817, demonstrating that the $Z^{2/3}/\sqrt{Q_\alpha}$ scaling coordinate provides a robust tool for modeling the radioactive engines of neutron-star mergers.
\end{abstract}

\maketitle

\section{Introduction}
Alpha decay provides one of the clearest manifestations of quantum tunneling in a finite, self-bound many-body system. Since the pioneering work of Geiger and Nuttall~\cite{geiger1911} and the quantum-mechanical explanation by Gamow and by Condon and Gurney~\cite{gamow1928,condon1928}, an exponential dependence of the decay half-life on the released energy has been firmly established. Nevertheless, extending this basic relation into a quantitative description valid across the nuclear chart has remained a long-standing challenge.

In the era of multimessenger gravitational-wave astrophysics, the macroscopic observables of neutron-star mergers, specifically the bolometric light curves and spectral evolution of kilonovae, are governed by the microscopic radioactive decay of thousands of freshly synthesized, highly unstable nuclei. However, the $\alpha$-decay properties of the heavy, neutron-rich actinides that dominate the late-time ($t \gtrsim 10$ days) heating remain largely unmeasured. Uncertainties in these nuclear lifetimes directly propagate into systematic biases when inferring ejecta masses, velocities, and opacities from gravitational-wave counterparts~\cite{wu2019,kasen2017}, with $\alpha$-decay and fission heating-rate prescriptions shown to alter predicted late-time luminosities by up to an order of magnitude depending on the adopted mass model~\cite{rosswog2017,zhu2021}.

Numerous phenomenological and semi-empirical relations, including the Viola--Seaborg formula~\cite{viola1966}, the Royer formula~\cite{royer2000}, and the Universal Decay Law (UDL)~\cite{qi2009}, reproduce experimental $\alpha$-decay half-lives with impressive accuracy. However, this success is achieved through the introduction of several fitted coefficients and, in some cases, nucleus-dependent hindrance factors. As a result, these formulations do not isolate a single scaling variable that captures the essential physics of $\alpha$ decay in a model-independent manner. This limitation becomes particularly acute when extrapolating to the neutron-rich, experimentally inaccessible regions populated by the rapid neutron-capture $r$-process in neutron-star merger ejecta.

Rather than constructing another parametrization, a complementary approach is to examine whether existing experimental data and theoretical descriptions share a hidden scaling structure when expressed in an appropriate variable. If such a variable exists, it should emerge consistently across distinct models and microscopic calculations, despite their different assumptions and functional forms. Here we investigate whether a universal scaling law of the form
\begin{equation}
\log_{10} T_{1/2} = m_a X_a + c_a, \qquad
X_a = \frac{Z^a}{\sqrt{Q_\alpha}}
\label{eq:scaling}
\end{equation}
can capture the essential physics of $\alpha$ decay. We demonstrate that both experiment and quantum-mechanical tunneling theory converge to a unique scaling variable with a specific value of the exponent $a$, revealing an emergent tunneling universality that transcends model-specific parametrizations. The resulting scaling relation establishes a semi-analytic link between the microscopic $\alpha$-decay systematics and the emergent quasi-power-law radioactive heating behavior of neutron-star merger ejecta, establishing a direct link between fundamental quantum tunneling and multimessenger astrophysical observables.

\section{Universal Scaling Law from Global Optimization}
We first examine the scaling behavior implied by Eq.~(\ref{eq:scaling}) using the complete set of measured $\alpha$-decay half-lives for even--even nuclei, spanning $^{106}$Te to $^{294}$Og and covering more than twenty orders of magnitude in lifetime~\cite{IAEA-LiveChart}. For each trial exponent $a$, we perform a linear regression of $\log_{10}T_{1/2}$ against $X_a$ and quantify the quality of the resulting scaling relation through the adjusted coefficient of determination, $R^2_{\mathrm{adj}}(a)$.

\begin{figure}[htbp]
\centering
\includegraphics[width=0.48\textwidth]{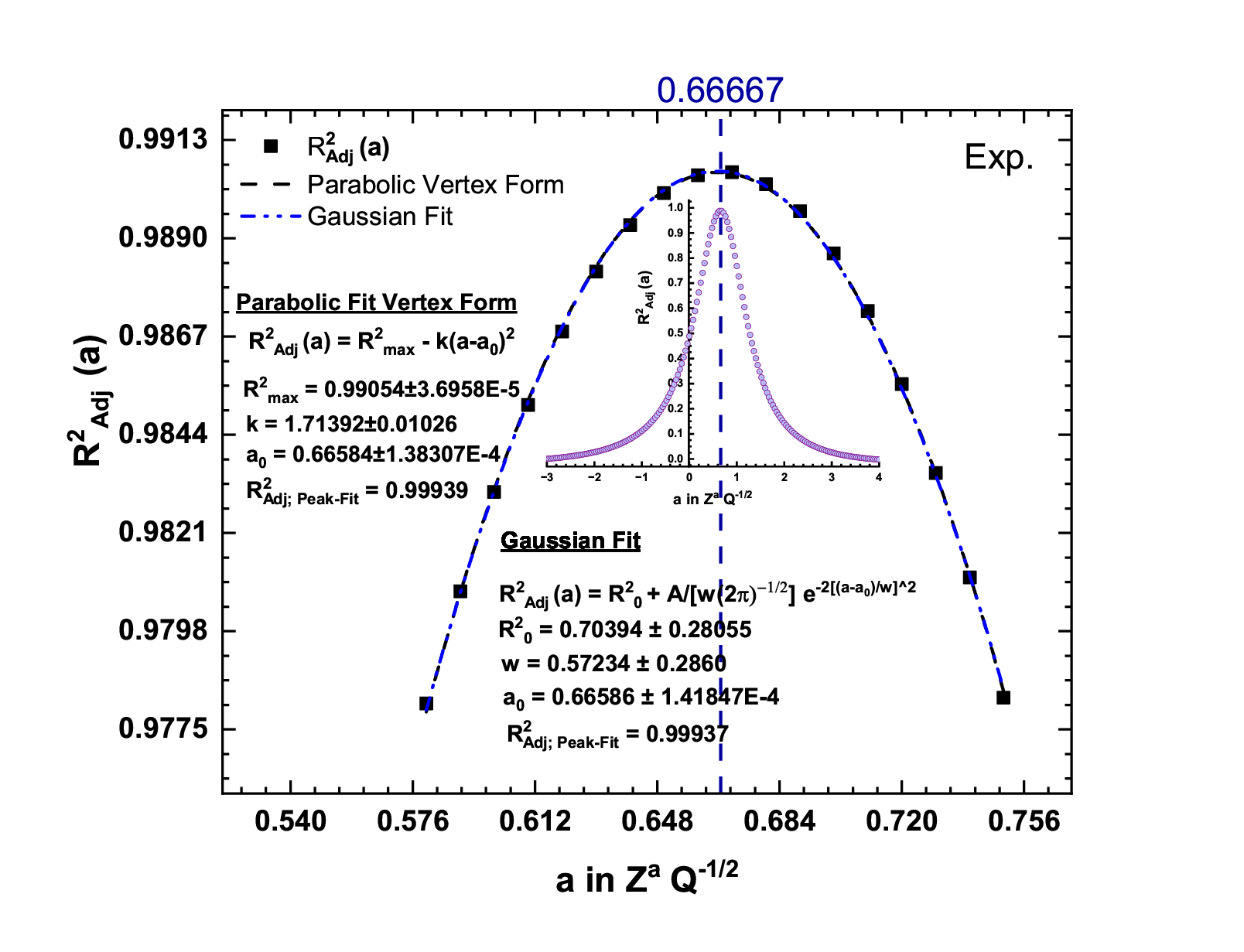}
\includegraphics[width=0.48\textwidth]{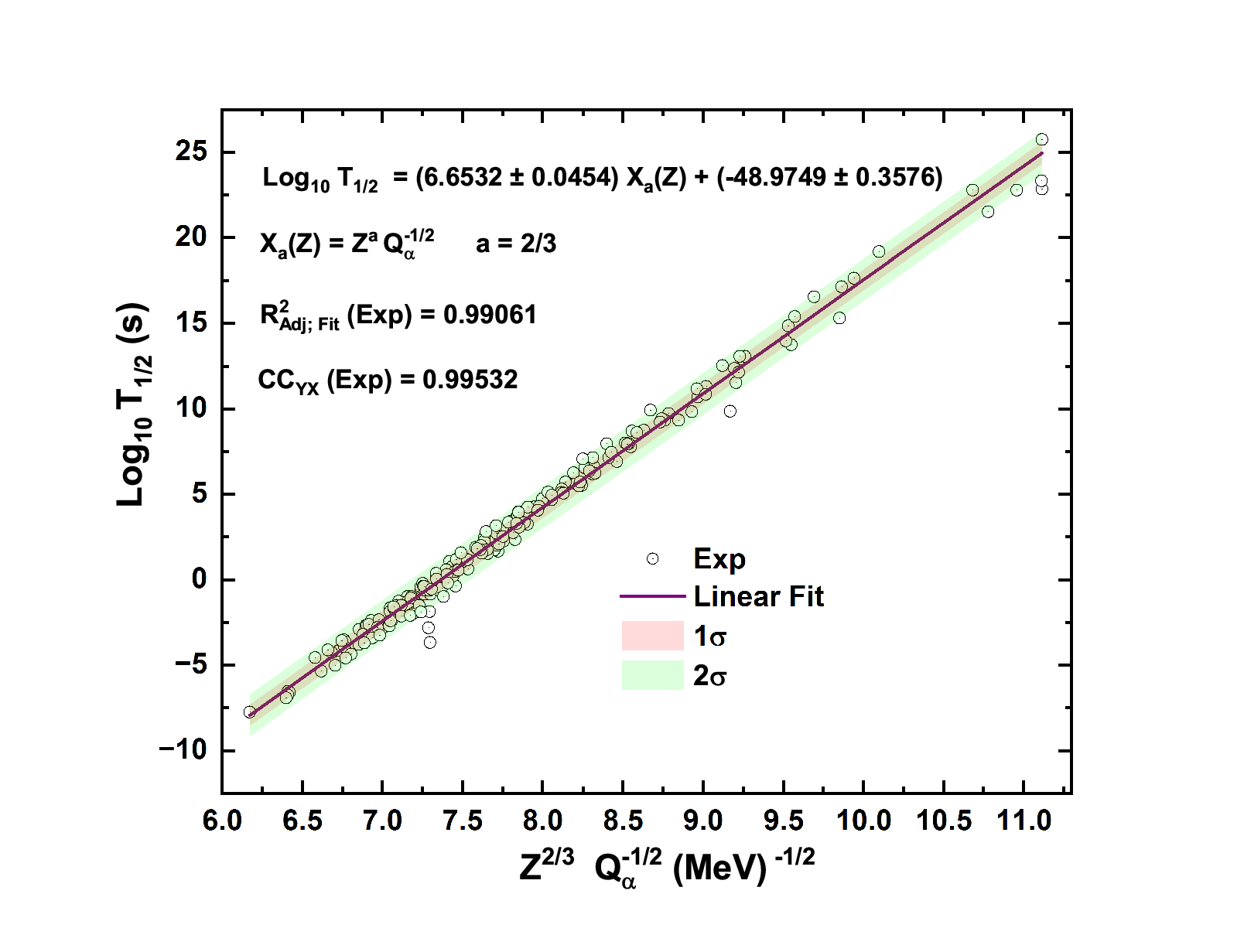}
\caption{(a) The adjusted coefficient of determination as a function of the exponent $a$ for the experimental even--even $\alpha$ emitters. The main panel shows the vicinity of the maximum, while the inset displays the extended range $a\in[-3,4]$. Both views reveal a single, sharply defined global maximum at $a\simeq 2/3$. (b) Experimental $\alpha$-decay half-lives plotted as a function of $Z^{2/3}/\sqrt{Q_\alpha}$. The linear correlation spans more than twenty orders of magnitude in lifetime, yielding the compact relation in Eq.~(\ref{eq:final}).}
\label{fig1}
\end{figure}

Figure~\ref{fig1}a reveals a striking feature: $R^2_{\mathrm{adj}}(a)$ exhibits a single, sharply defined global maximum at $a_0 = 0.66584 \pm 0.00014$, with no competing local extrema across the extensive range $a \in [-3,4]$. Both parabolic and Gaussian fits to the peak yield statistically indistinguishable values ($a_0 = 0.66584$--$0.66586$), confirming the robustness of the optimum. Critically, fixing the exponent to the canonical value $a = 2/3$ produces a correlation that is statistically indistinguishable from the global optimum ($\Delta R^2_{\mathrm{adj}} < 2\times10^{-4}$), while the conventional Gamow scaling ($a=1$) yields a substantially inferior fit ($\Delta R^2_{\mathrm{adj}} \approx 0.22$).

The sharpness and uniqueness of this optimum indicate that $a \simeq 2/3$ reflects a nontrivial scaling structure inherent to heavy $\alpha$ decay. A natural theoretical motivation for this exponent follows from the structure of conventional $\alpha$-decay systematics, which contain both a Coulomb-barrier contribution, scaling approximately as $Z/\sqrt{Q_\alpha}$, and a $Q_\alpha$-independent geometric contribution involving $A^{1/6}Z^{1/2}$. In the heavy-nucleus region of the nuclear chart, the mass-to-charge ratio varies relatively slowly, with $A/Z$ remaining of order $2.5$, so that $A \simeq CZ$ provides a useful leading-order approximation, with $C$ varying only moderately across the region considered. The geometric contribution then scales as $A^{1/6}Z^{1/2} \simeq C^{1/6}Z^{2/3}$, naturally identifying $2/3$ as the characteristic charge exponent associated with the nuclear-size contribution. The experimentally determined optimum is therefore theoretically motivated by the leading geometric scaling of heavy nuclei, while the accompanying $Q_\alpha^{-1/2}$ dependence reflects the dominant Coulomb-barrier sensitivity. Rather than constituting an exact algebraic identity, the resulting variable $X_{2/3}=Z^{2/3}/\sqrt{Q_\alpha}$ should be understood as an emergent one-dimensional representation of these correlated nuclear dependencies, whose quantitative validity is established by the global optimization and independently supported by distinct theoretical descriptions.

Evaluating the experimental scaling at the canonical exponent $a = 2/3$ yields the compact relation
\begin{equation}
\log_{10} T_{1/2} = (6.653 \pm 0.045)\,\frac{Z^{2/3}}{\sqrt{Q_\alpha}} - (48.975 \pm 0.358)
\label{eq:final}
\end{equation}
which collapses all data points onto a single line with $R^2_{\mathrm{adj, Fit}} = 0.99061$ (Fig.~\ref{fig1}b). In this relation, the slope ($6.653 \pm 0.045$) encapsulates the combined sensitivity of the Coulomb barrier penetration and the nuclear surface preformation, while the intercept ($-48.975 \pm 0.358$) absorbs the assault frequency and intrinsic preformation constants. The linear correlation spans more than twenty orders of magnitude in lifetime, with a symmetric residual distribution characterized by a standard deviation of $\sigma = 0.50$ dex. This remarkable precision demonstrates that $Z^{2/3}/\sqrt{Q_\alpha}$ captures the essential physics of $\alpha$ decay across the entire nuclear chart, reducing a complex many-body tunneling problem to a single, model-independent scaling variable.

\section{Convergence of Independent Theoretical Models}
The universality of the scaling is further tested against six established $\alpha$-decay approaches that span a wide range of physical assumptions and functional forms. These include our independent microscopic WKB calculations employing double-folding $\alpha$-nucleus potentials with microscopically computed preformation factors~\cite{cpc2026}, together with five widely used phenomenological and semi-empirical formulations: the Royer liquid-drop formula~\cite{royer2000}, the Universal Decay Law (UDL)~\cite{qi2009}, the New Geiger--Nuttall formula (NGNF)~\cite{ren2012}, the Denisov--Khudenko model~\cite{denisov2009}, and the Brown formula~\cite{brown1992}. 

\begin{figure}[htbp]
\centering
\includegraphics[width=0.48\textwidth]{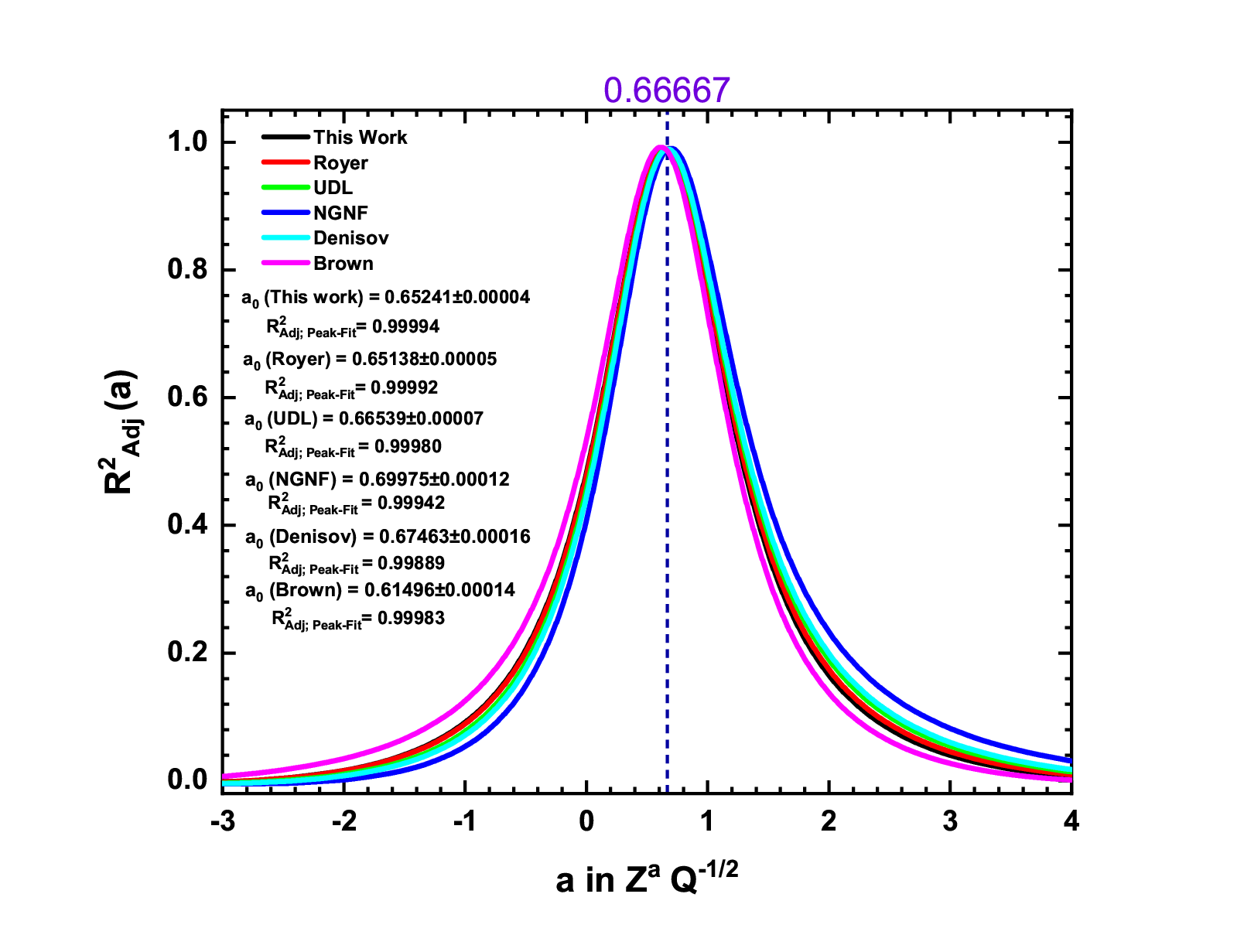}
\includegraphics[width=0.48\textwidth]{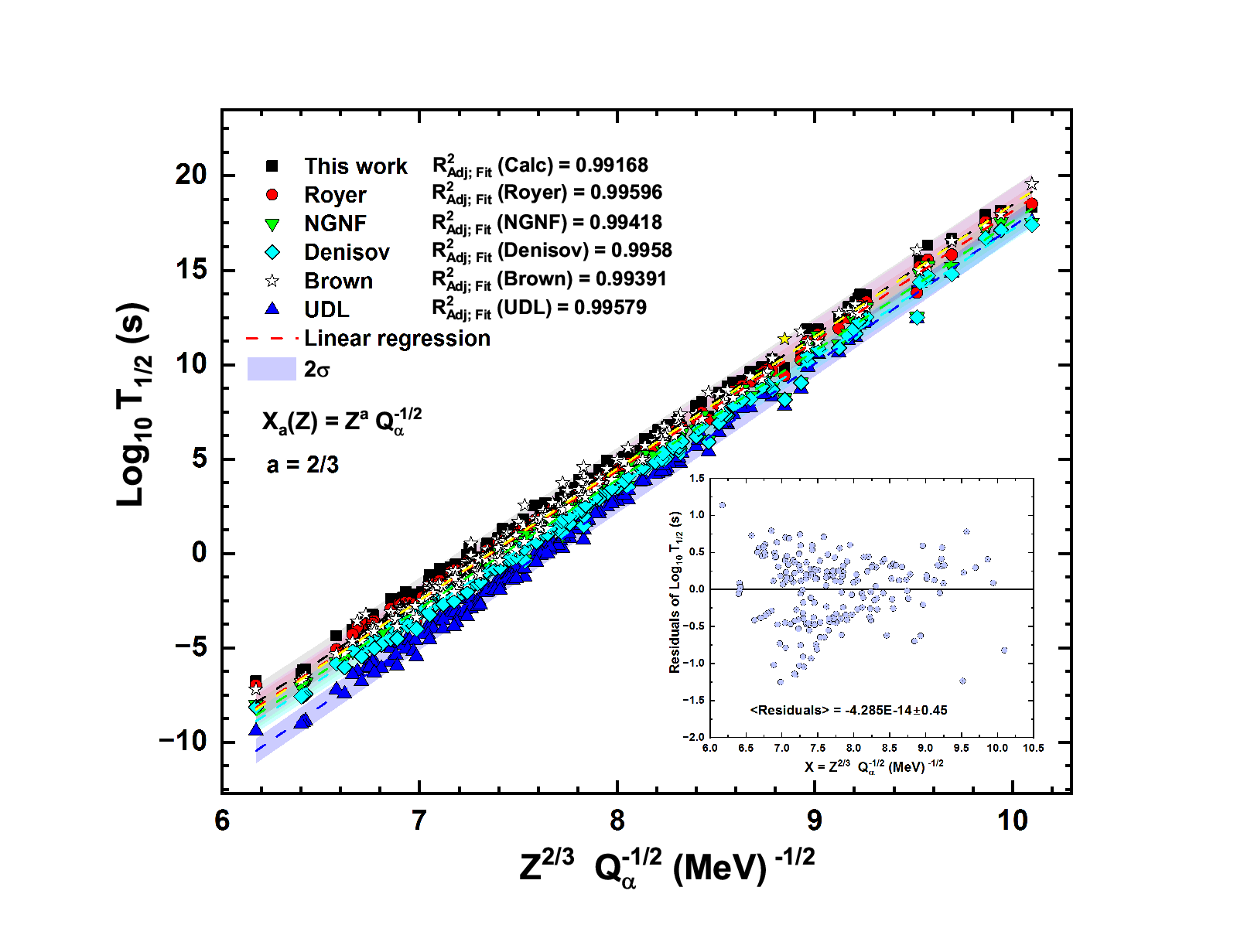}
\caption{(a) The $R^2_{\mathrm{adj}}(a)$ for the six theoretical models across the full scanned range, demonstrating that each model independently exhibits a single, unimodal peak clustered tightly around $2/3$. (b) Theoretical predictions from the six approaches plotted against $Z^{2/3}/\sqrt{Q_\alpha}$ using the fixed exponent $a=2/3$. All models collapse onto linear trends with $R^2_{\mathrm{adj, Fit}} > 0.99$ using their original parameter sets without refitting. The inset shows the WKB residual distribution, symmetric about zero.}
\label{fig2}
\end{figure}

When each approach is independently optimized by varying the exponent $a$ in $X_a = Z^a/\sqrt{Q_\alpha}$, the resulting optima reveal a profound structural agreement (Fig.~\ref{fig2}a). For the experimental data, the vertex of the parabolic interpolation of the $R^2_{\rm adj}(a)$ curve occurs at $a_0=0.66584\pm0.00014$, with a peak $R^2_{\mathrm{adj, Peak-Fit}}=0.99939$. Remarkably, our independent microscopic WKB calculations yield a comparably sharp optimum at $a_0=0.65241\pm0.00004$, with $R^2_{\mathrm{adj, Peak-Fit}}=0.99994$. The convergence persists across the semi-empirical descriptions: the Royer, UDL, and Denisov--Khudenko formulations independently optimize at $a_0=0.65138\pm0.00005$, $0.66539\pm0.00007$, and $0.67463\pm0.00016$, respectively, all clustering within a narrow interval around $2/3$, with peak fit qualities exceeding $0.998$. The NGNF formulation gives $a_0=0.69975\pm0.00012$, while the Brown relation yields $a_0=0.61496\pm0.00014$, reflecting the distinct functional structure of those parametrizations, such as the explicit $(Z-2)^{0.6}$ dependence in the latter. The unweighted mean of all six independently determined exponents is $a_0=0.660$, with a model-to-model standard deviation of $0.028$, placing the collective optimum within $1\%$ of $2/3$. 

Consequently, fixing the exponent at the canonical value $a=2/3$ provides a direct test of whether this independently optimized convergence translates into a common linear scaling without the need to refit the underlying model parameters. As shown in Fig.~\ref{fig2}b, all six approaches exhibit highly linear correlations between $\log_{10}T_{1/2}$ and $X_{2/3}=Z^{2/3}/\sqrt{Q_\alpha}$, retaining $R^2_{\mathrm{adj, Fit}}>0.99$ using their original parameter sets. As an independent microscopic calculation, the WKB results provide the most direct theoretical realization of this scaling and give
\begin{equation}
\log_{10}T_{1/2}
=(6.887\pm0.042)\frac{Z^{2/3}}{\sqrt{Q_\alpha}}
-(50.390\pm0.324),
\label{eq:wkb}
\end{equation}
with $R^2_{\mathrm{adj, Fit}}=0.99281$. The five semi-empirical descriptions likewise produce exceptionally linear correlations, with $R^2_{\mathrm{adj, Fit}}$ ranging from $0.99382$ to $0.99678$ and fitted slopes spanning $6.80$--$7.25$. The inset of Fig.~\ref{fig2}b shows the residual distribution for the WKB correlation, which is approximately symmetric about zero with a standard deviation of $\sigma=0.50$ dex. Taken together, the independently optimized exponents, their collective concentration near $2/3$, and the persistence of the fixed-$a$ linear correlations provide strong evidence that $Z^{2/3}/\sqrt{Q_\alpha}$ is an emergent scaling coordinate shared by otherwise distinct descriptions of the dominant $\alpha$-decay systematics.

\section{Implications for Multimessenger Kilonova Observables}

The universal scaling provides a minimal two-parameter framework for extrapolating $\alpha$-decay lifetimes into the neutron-rich regions populated by the rapid neutron-capture ($r$) process in binary neutron-star mergers. The multimessenger observation of the gravitational-wave event GW170817~\cite{abbott2017} and its electromagnetic counterpart, the kilonova AT2017gfo~\cite{kasen2017}, demonstrated that the macroscopic bolometric light curves of these events are powered by the radioactive decay of freshly synthesized nuclei.

While $\beta$-decay dominates the early-time heating ($t \lesssim 1$ week) from short-lived isotopes, the $\alpha$-decay of heavy nuclei becomes the primary heat source at late times ($t \gtrsim 10$ days). Because the $\alpha$-decay properties of these heavy, neutron-rich nuclei remain largely unmeasured, kilonova models rely on theoretical mass models and phenomenological extrapolations. The resulting nuclear uncertainties propagate directly into the inferred ejecta properties, such as mass, velocity, and opacity, extracted from multimessenger observations.

Establishing a rigorous connection between fundamental nuclear microphysics and multimessenger observables is a central objective in modern gravitational-wave astrophysics~\cite{prd2026, metzger2017, wu2023}. In the context of neutron-star mergers, the late-time electromagnetic emission is highly sensitive to the decay properties of exotic, neutron-rich isotopes that currently lack experimental constraints. We address this challenge by propagating the microscopic scaling relation into the macroscopic radioactive heating through an explicit numerical treatment of the underlying $\alpha$-decay population. By propagating these predictions through detailed radiative-diffusion calculations, we model the observable kilonova light curves, thereby providing a robust constraint on late-time heating that circumvents the systematic uncertainties of traditional nuclear mass models.

\subsection{Analytic Derivation of Power-Law Heating}
The macroscopic radioactive heating rate in merger ejecta is a critical input for modeling kilonova light curves. It is well established that the total heating rate from a complex mixture of decaying nuclei follows a quasi-power-law behavior, $\dot{\epsilon} \propto t^{-\alpha}$, with $\alpha \approx 1.2$--$1.4$~\cite{metzger2010,korobkin2012}. This behavior arises because the distribution of decay constants in the $r$-process abundance pattern is roughly uniform in $\ln \lambda$. 

Substituting the scaling law into the definition of the decay constant yields $\ln\lambda = \ln(\ln 2) - \ln(10)(m X_{2/3} + c)$. Because $\ln\lambda$ is strictly linear with respect to $X_{2/3}$, the requirement for power-law heating reduces to a directly testable statement: if the scaling variable $X_{2/3}$ is roughly uniformly distributed across the population of decaying nuclei, the decay constants $\ln\lambda$ will share this uniformity, and the macroscopic heating rate will naturally follow a power law.

\begin{figure}[htbp]
\centering
\includegraphics[width=0.48\textwidth]{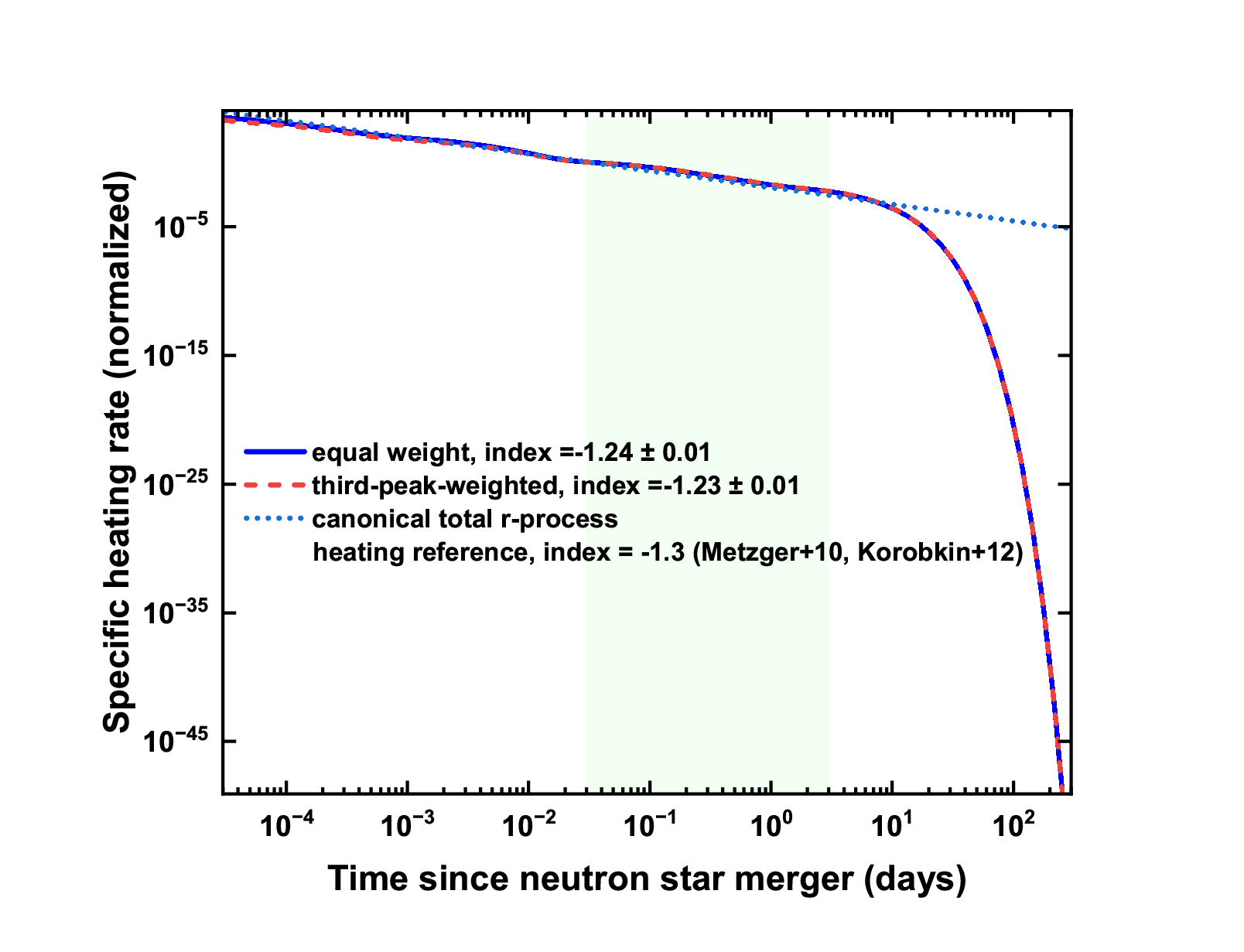}
\caption{Numerically summed $\alpha$-decay heating rate for 36 experimentally characterized even--even $\alpha$-emitters spanning the heavy-mass region ($Z=70$--$84$, $Q_\alpha \approx 5.5$--$7.5$ MeV), under two abundance-weighting assumptions (equal weight and third-$r$-process-peak-weighted), with a power-law fit over the physically motivated window $t=0.03$--$3$~days (shaded). Both weightings give indices within $0.01$ of each other and of the commonly adopted total $r$-process heating reference index $-1.3$~\cite{metzger2010,korobkin2012}, shown for shape comparison.}
\label{fig:heating}
\end{figure}

We examine this directly using 36 experimentally characterized even-even \(\alpha\)-emitters spanning the heavy-mass region ($Z=70$--$84$) with $Q_\alpha$ values in the range $5.5$--$7.5$ MeV. This region is astrophysically motivated by the role of heavy $\alpha$-decaying nuclei in the radioactive heating of neutron-star merger ejecta~\cite{mumpower2018,wu2023}. A Kolmogorov--Smirnov test of their $X_{2/3}$ values against a uniform distribution over the prescribed range yields $D=0.16$, $p=0.28$, consistent with uniformity. Summing $\dot{\epsilon}_\alpha(t) = \sum_i Y_i Q_{\alpha,i} \lambda_i e^{-\lambda_i t}$ over this population, taking equal per-nucleus weights for this diagnostic sum, and fitting a local power law over the interval $t=0.03$--$3$ days, where the summed curve remains approximately power-law before appreciable discrete-nucleus deviations develop, gives a numerical index of $-1.24 \pm 0.01$ (Fig.~\ref{fig:heating}). This index is close to the $t^{-1}$ leading-order expectation and the commonly adopted $\beta$-decay heating index of approximately $-1.3$. This provides a semi-analytic connection between the microscopic scaling law and the macroscopic quasi-power-law heating behavior, supported by the statistical consistency of the $X_{2/3}$ distribution with uniformity rather than imposed through an empirical heating ansatz.

\subsection{Absolute Thermalization and Heating Rates}
To validate this analytically derived index against absolute dimensional heating, we perform independent numerical calculations of the thermalized heating rates. The instantaneous nuclear heating rate is given by $\dot{\epsilon}_{\rm nuc}(t) = \sum_i Y_i Q_{\alpha,i} \lambda_i e^{-\lambda_i t}$. In the expanding ejecta, the observable thermalized heating rate is $\dot{\epsilon}_{\rm th}(t) = \dot{\epsilon}_{\rm nuc}(t) f_{\rm th}(t)$, where $f_{\rm th}(t)$ is the thermalization efficiency function. As established in foundational works on kilonova thermalization~\cite{barnes2016,barnes2019,hotokezaka2016}, $f_{\rm th}(t)$ is governed by the ejecta density evolution $\rho(t) \propto t^{-3}$ and the energy-dependent stopping cross-sections for $\gamma$-rays, positrons, and $\alpha$-particles. While $f_{\rm th} \approx 1$ at early times, it declines at late epochs as the ejecta becomes optically thin to $\gamma$-rays.

\begin{figure}[htbp]
\centering
\includegraphics[width=0.49\textwidth]{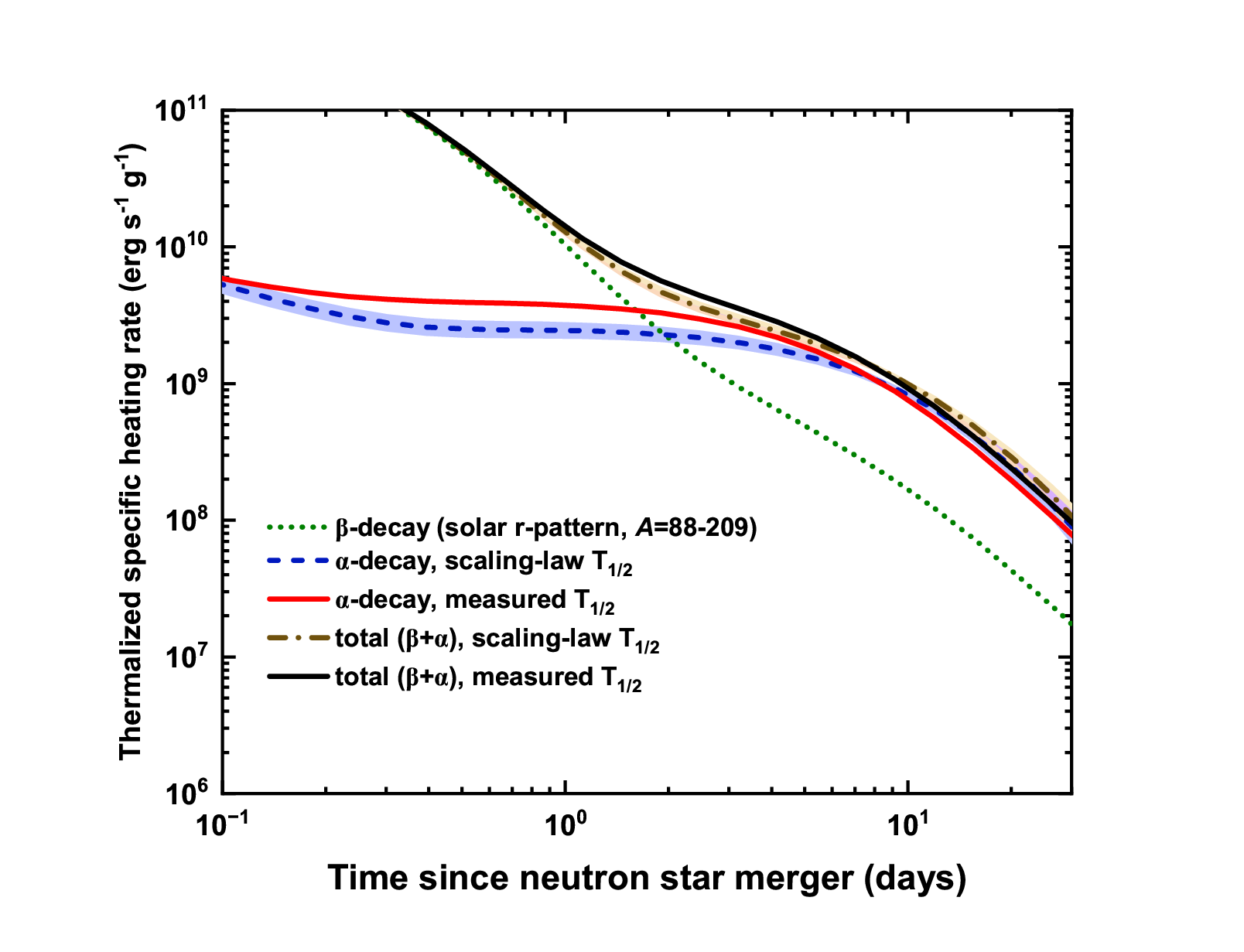}
\includegraphics[width=0.49\textwidth]{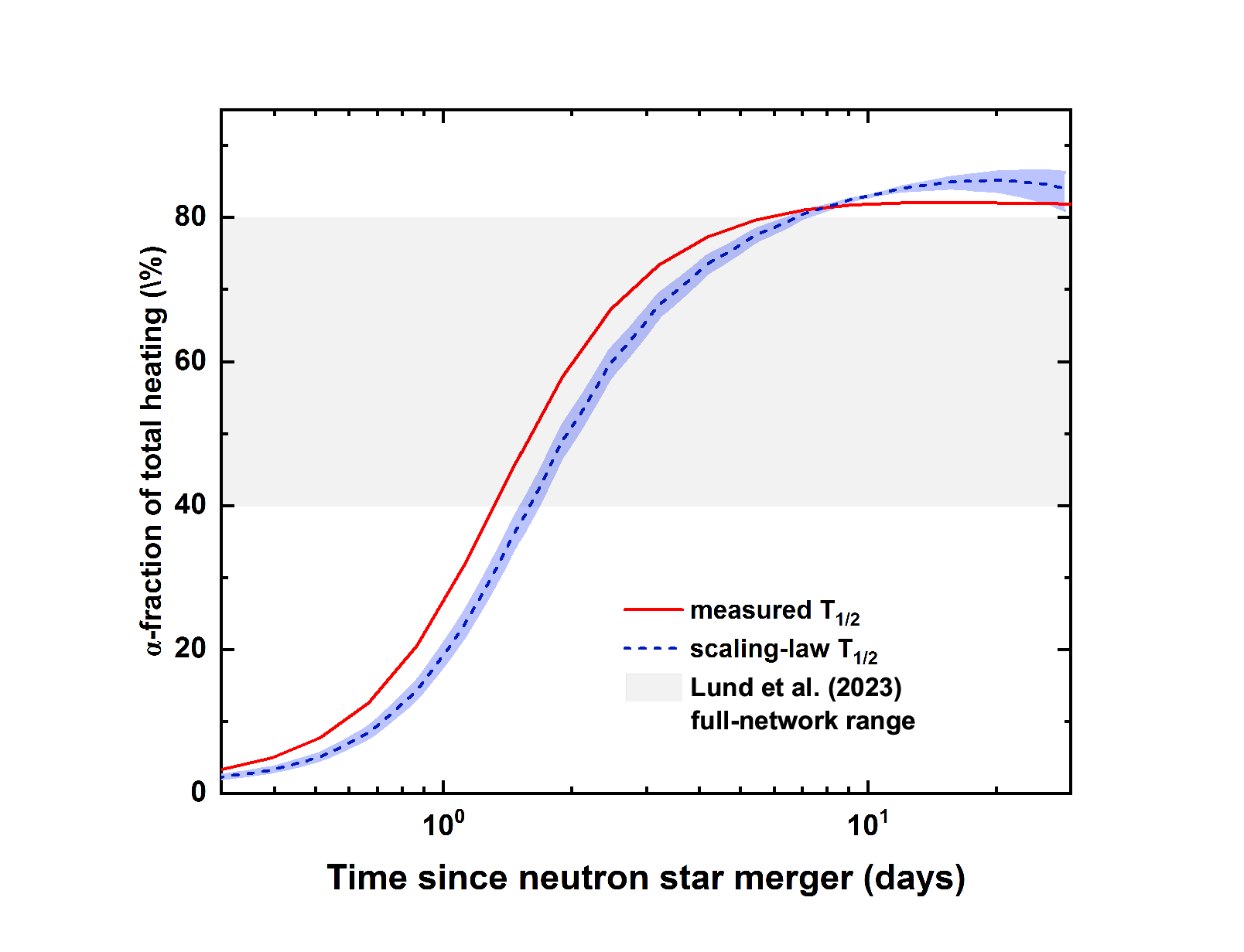}
\caption{(a) The simulated thermalized specific heating rates using real ENDF/B-VII.1 injection spectra and full thermalization physics, applied as published and with the two in-domain chain half-lives replaced by our scaling-law predictions. (b) The resulting, fully self-consistent fractional $\alpha$-contribution (Table~\ref{tab:heating_compare}) against the Lund et al.~\cite{lund2023} full-network benchmark.}
\label{fig:realheating}
\end{figure}

We integrate the $\alpha$-decay half-lives predicted by our universal scaling law directly into the nuclear source terms for the four dominant bottleneck chains identified by Wu et al.~\cite{wu2019} ($^{222}$Rn, $^{223}$Ra, $^{224}$Ra, $^{225}$Ra$\to^{225}$Ac). Replacing the measured half-lives of the two in-domain even--even chains ($^{222}$Rn, $^{224}$Ra) with our scaling-law predictions allows us to test the macroscopic impact of the scaling-law inputs. The uncertainties reported for the scaling-law column are derived from $10^5$ correlated Monte Carlo realizations of the fit covariance in Eq.~(\ref{eq:final}) incorporating the strong anti-correlation ($\rho=-0.99$) to rigorously capture the non-linear error propagation into the heating rates. As shown in Table~\ref{tab:heating_compare} and Fig.~\ref{fig:realheating}, the simulated $\alpha$-decay fractional contribution to total heating reaches a plateau of $\approx 80$--$85\%$ at $t \gtrsim 7$ days when using scaling-law inputs, compared to $\approx 82\%$ when using measured half-lives. Both calculations remain fully consistent with independent full-network benchmarks~\cite{lund2023}. This confirms that the $Z^{2/3}/\sqrt{Q_\alpha}$ scaling variable reliably captures the aggregate $\alpha$-decay physics required for kilonova modeling, even when individual half-life predictions carry inherent uncertainties. We note that spontaneous fission (e.g., of $^{254}$Cf, $T_{1/2}\approx60$~days) provides a complementary late-time heating channel outside the scope of this $\alpha$-decay-focused analysis~\cite{zhu2018}.

\begin{table}[htbp]
\centering
\caption{Fractional $\alpha$-decay contribution to the total ($\beta + \alpha$) heating rate in neutron-star merger ejecta. The values are computed using a self-consistent thermalization treatment for the dominant bottleneck chains ($^{222}$Rn, $^{223}$Ra, $^{224}$Ra, and $^{225}$Ra$\to^{225}$Ac). We compare the results obtained using experimental half-lives versus those obtained by replacing the in-domain even--even chains, $^{222}$Rn and $^{224}$Ra, with the universal scaling-law predictions.}
\label{tab:heating_compare}
\begin{tabular}{@{}ccc@{}}
\hline
$t$ (days) \hspace*{0.3cm} & Measured $T_{1/2}$ \hspace*{0.3cm} & Scaling-Law $T_{1/2}$ \\ 
\hline 
1   & 32.0\% & \textbf{$19.3^{+1.6}_{-1.5}$\%} \\
2   & 57.9\% & \textbf{$51.0^{+2.2}_{-2.2}$\%} \\
4   & 77.3\% & \textbf{$72.6^{+1.2}_{-1.4}$\%} \\
7   & 81.1\% & \textbf{$80.4^{+0.4}_{-0.6}$\%} \\
15  & 82.1\% & \textbf{$84.8^{+0.6}_{-0.8}$\%} \\
30  & 82.0\% & \textbf{$83.8^{+2.4}_{-3.3}$\%} \\
60  & 80.2\% & \textbf{$73.2^{+8.0}_{-11.5}$\%} \\
\hline
\end{tabular}
\end{table}

\subsection{Bolometric Light Curves and Spectral Evolution}
The fractional-heating comparison is downstream of a radiative-transfer step: photon diffusion smooths and delays the observable luminosity relative to the instantaneous heating rate. We complete our analysis by propagating the calculated heating curves through a semi-analytic radiative-diffusion light-curve formalism~\cite{hotokezaka2020}. The bolometric luminosity is computed via the Arnett-like one-zone diffusion integral~\cite{arnett1982,metzger2017}:
\begin{equation}
L(t) \approx M_{\rm ej} e^{-(t/\tau_d)^2}
\int_0^t \frac{2t'}{\tau_d^2}
\dot{\epsilon}_{\rm th}(t') e^{(t'/\tau_d)^2} dt',
\end{equation}
where $\tau_d = \sqrt{2 \kappa M_{\rm ej} / (\beta c v_{\rm ej})}$ is the photon diffusion timescale, with $\beta \approx 13.7$ being a geometric constant for a uniform density sphere.

\begin{figure}[htbp]
\centering
\includegraphics[width=0.49\textwidth]{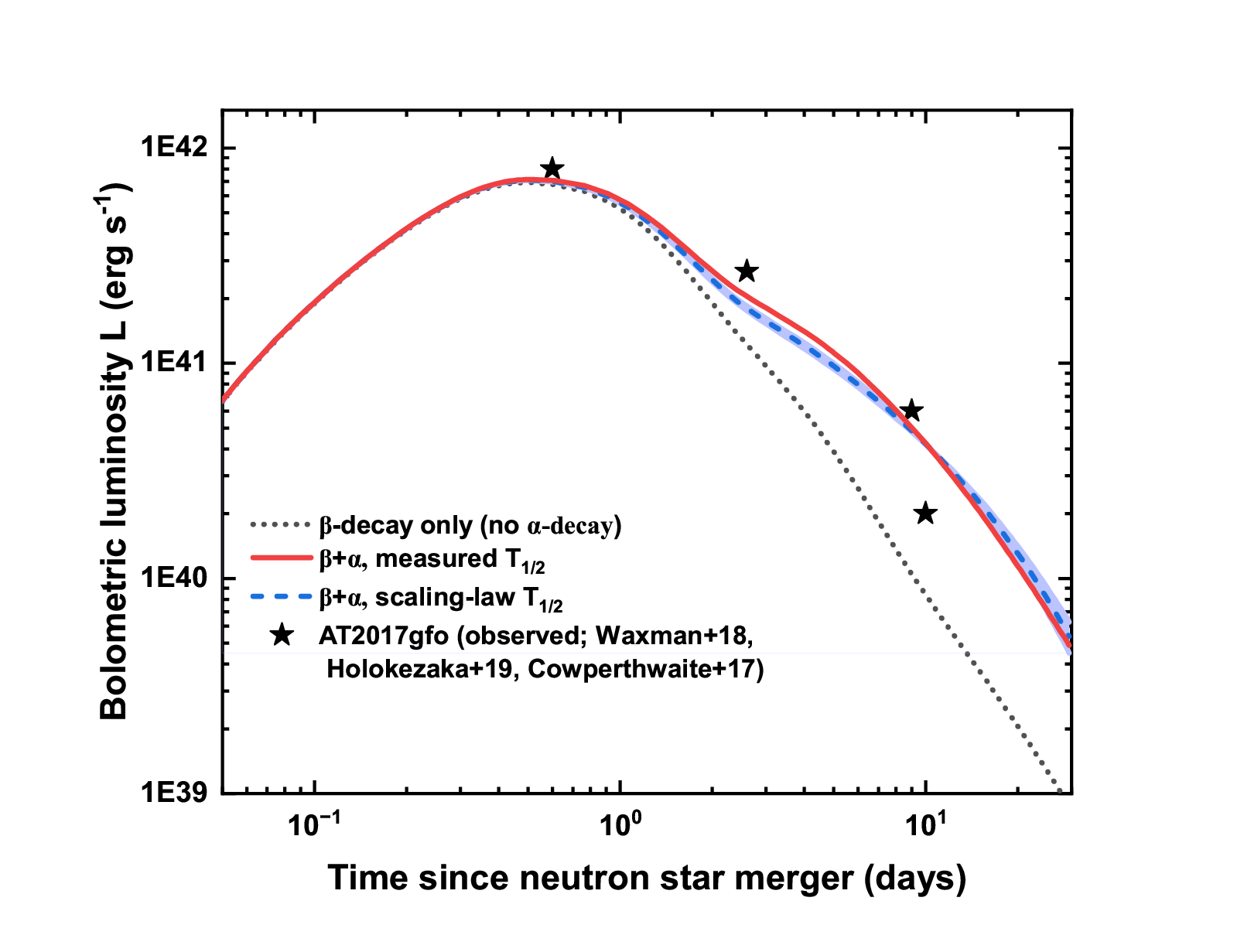}
\hfill
\includegraphics[width=0.49\textwidth]{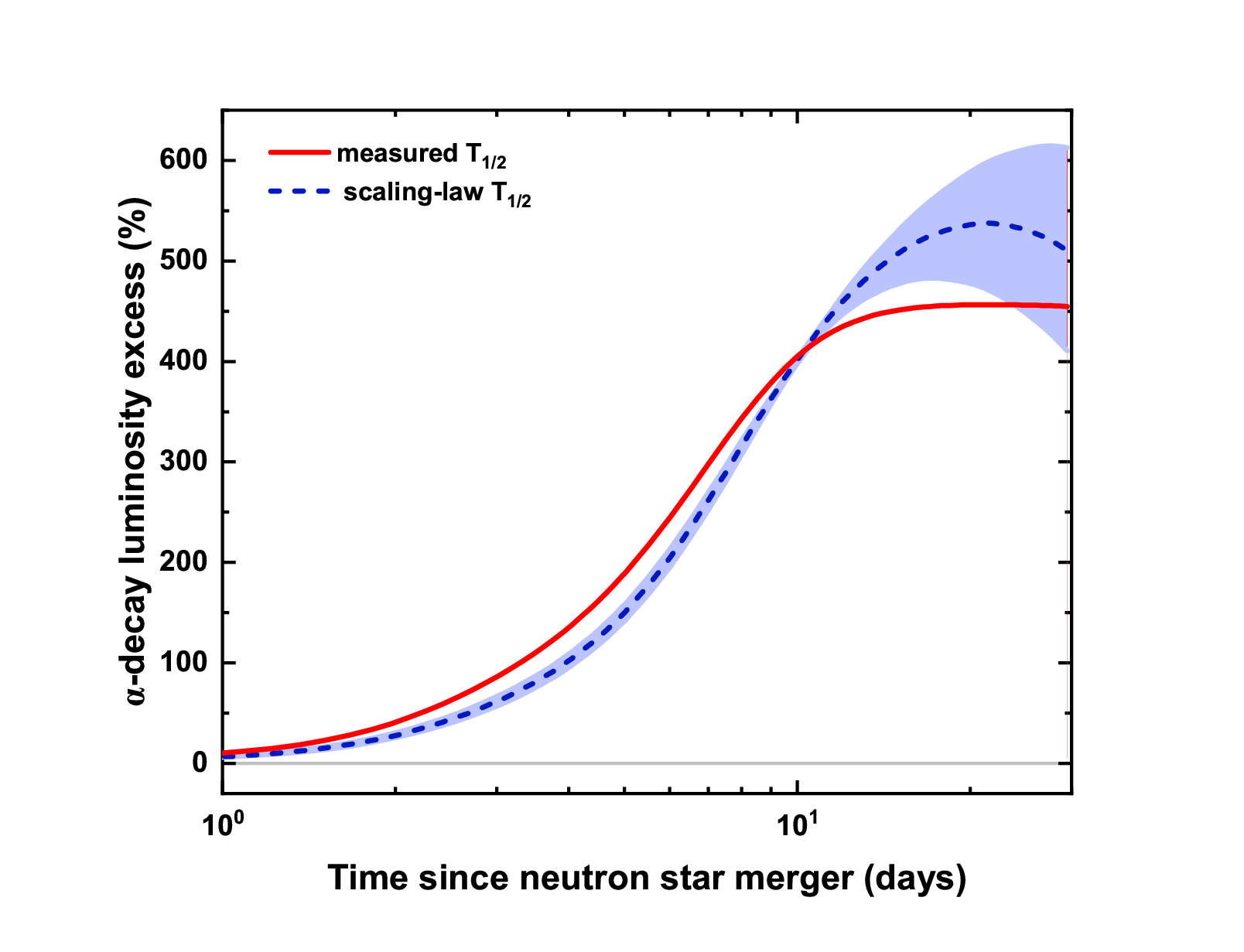}
\caption{(a) The resulting bolometric light curves from the radiative-diffusion formalism: $\beta$-decay only (dotted), measured- and scaling-law-input $\beta+\alpha$ (solid/dashed), against four real AT2017gfo observational anchors~\cite{waxman2018,cowperthwaite2017,fujibayashi2020} spanning peak to 10~days. (b) Fractional brightening due to $\alpha$-decay, up to $\approx 450\%$ around 15--20~days. The ratio of the two $\beta+\alpha$ model curves remains within $\pm 20\%$ throughout.}
\label{fig:lightcurve}
\end{figure}

The resulting bolometric light curves (Fig.~\ref{fig:lightcurve}) peak at $7.2 \times 10^{41}$ (measured input) and $7.1 \times 10^{41}$ (scaling-law input) erg s$^{-1}$ at $t \approx 0.5$ day. Both curves agree at peak because the peak is set by $\beta$-decay, before the $\alpha$-decay chains under study become important, and both track the independently observed AT2017gfo peak of $\approx 8 \times 10^{41}$ erg s$^{-1}$ at $t \approx 0.6$ day~\cite{waxman2018}.

Crucially, despite the residual microscopic scatter in the underlying half-lives, the derived scaling-law-input and measured-input light curves agree to within $\pm 20\%$ at every epoch from 1 to 29 days. The inclusion of $\alpha$-decay brightens the light curve by up to $\approx 450\%$ around 15--20 days relative to $\beta$-decay alone. Radiative diffusion further damps the fractional-heating-level discrepancies into a substantially smaller effect on the actual observable, an important and non-obvious conclusion that emerges only from running the full light-curve calculation.

\begin{figure}[htbp]
\centering
\includegraphics[width=0.49\textwidth]{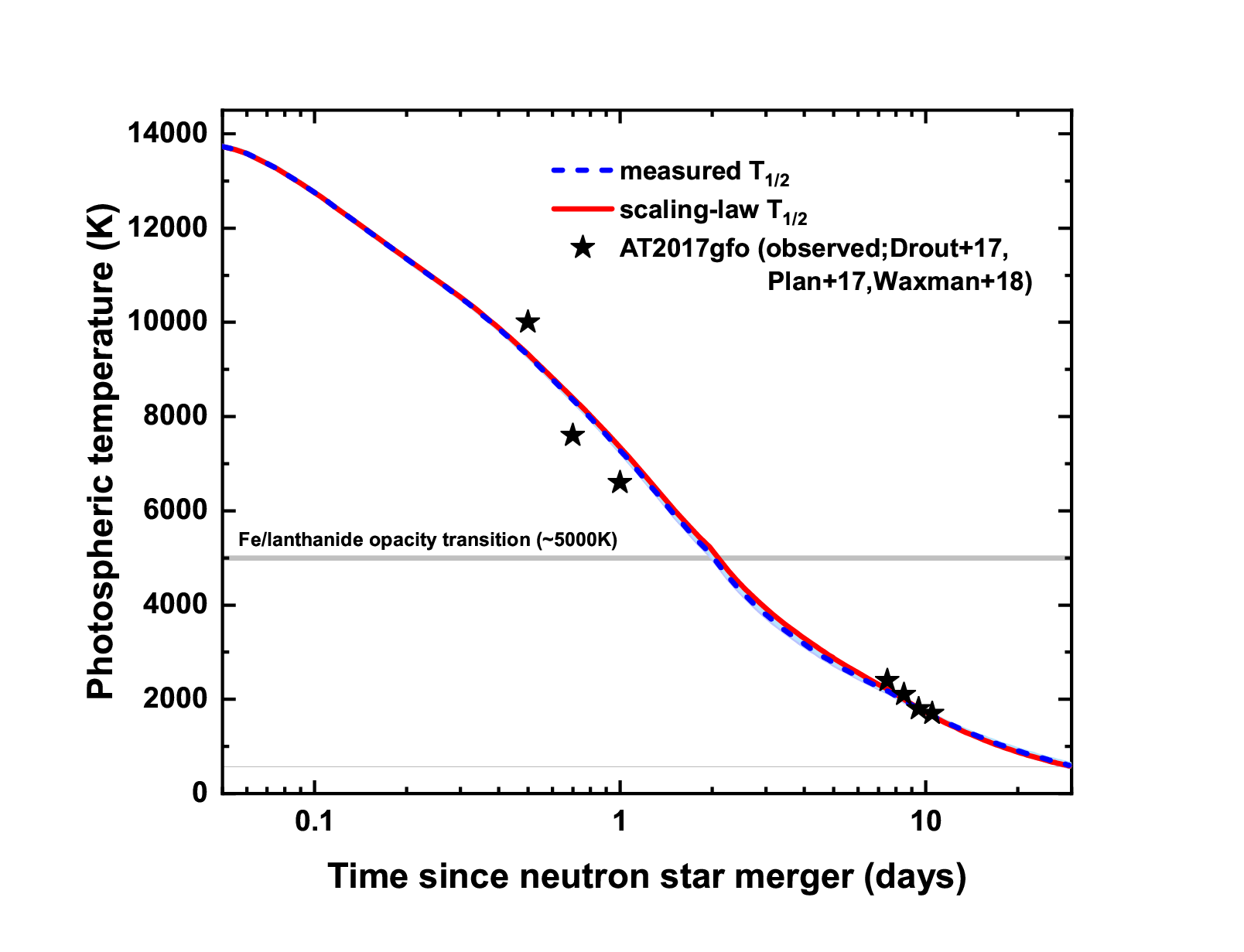}
\hfill
\includegraphics[width=0.49\textwidth]{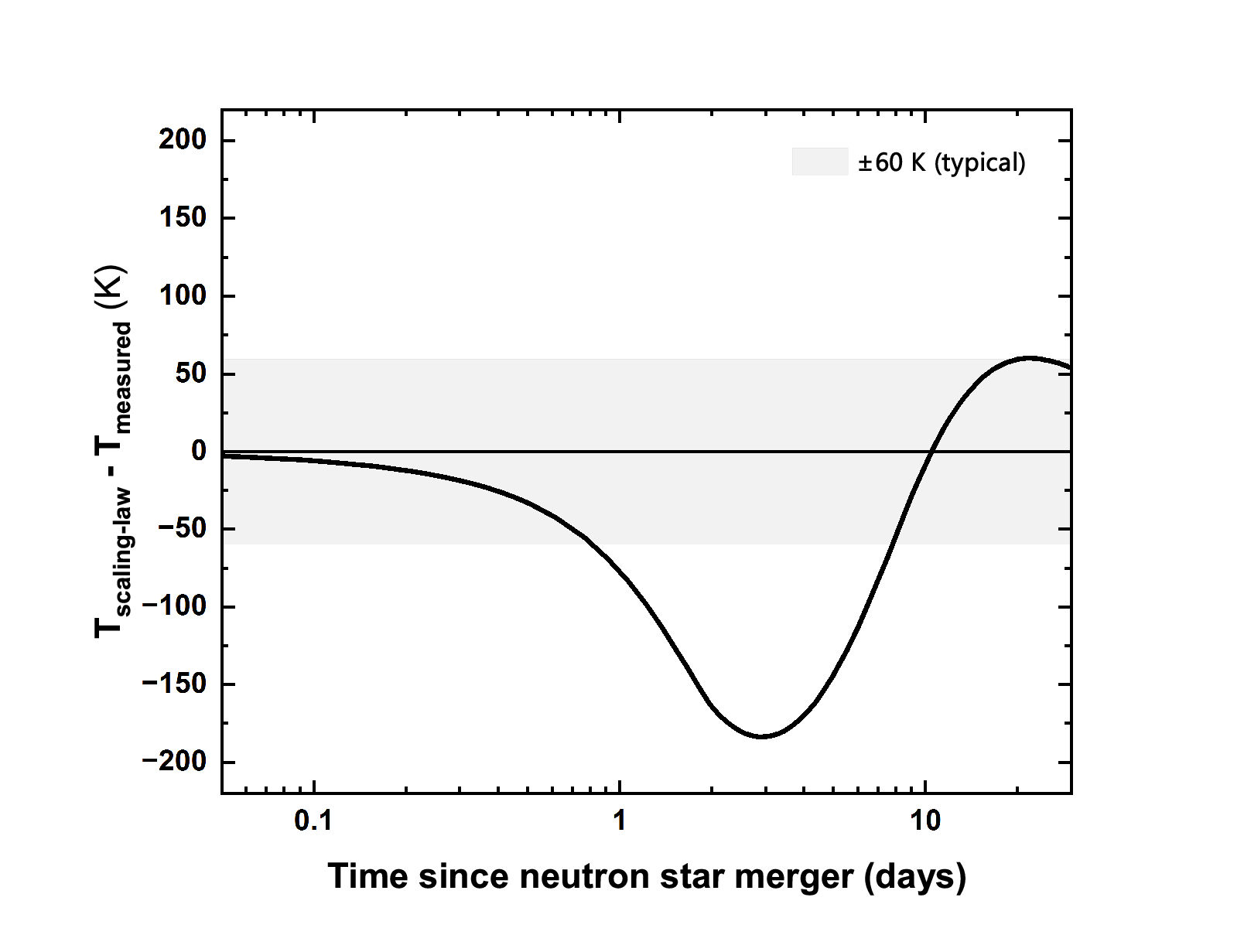}
\caption{(a) The effective blackbody temperature evolution for the measured- vs.\ scaling-law-input models, against seven real AT2017gfo blackbody temperature measurements spanning 0.5--10.5~days~\cite{drout2017,pian2017,waxman2018}. (b) The temperature difference between the two models: median $|\Delta T|=24$~K, maximum $136$~K, across the full range.}
\label{fig:temperature}
\end{figure}

The same calculation returns the effective blackbody temperature $T(t)$, derived from the Stefan-Boltzmann law $L = 4\pi R_{\rm ph}^2 \sigma T^4$ with $R_{\rm ph}(t) = v_{\rm ej} t$. As shown in Fig.~\ref{fig:temperature}, the temperature evolution is essentially indistinguishable between the measured- and scaling-law-input models, with a maximum difference of only 136 K across the full 0.05--30 day range. Both models track the observed cooling trend of AT2017gfo at the order-of-magnitude level, from $\sim 10^4$ K near peak down to $\sim 1700$ K in the nebular phase.

\subsection{Robustness Across Ejecta Geometries}
Neutron-star merger ejecta are inherently anisotropic, characterized by a fast, low-opacity polar wind and a slow, high-opacity equatorial dynamical component~\cite{villar2017,cowperthwaite2017}. The opacity $\kappa$ in merger ejecta is dominated by bound-bound transitions; lanthanide- and actinide-rich material exhibits opacities orders of magnitude higher ($\kappa \sim 10$ cm$^2$ g$^{-1}$) than iron-group elements due to the high density of $f$-shell energy levels, driving the characteristic red/infrared emission of kilonovae~\cite{kasen2013,tanaka2013}. 

\begin{figure}[htbp]
\centering
\includegraphics[width=0.49\textwidth]{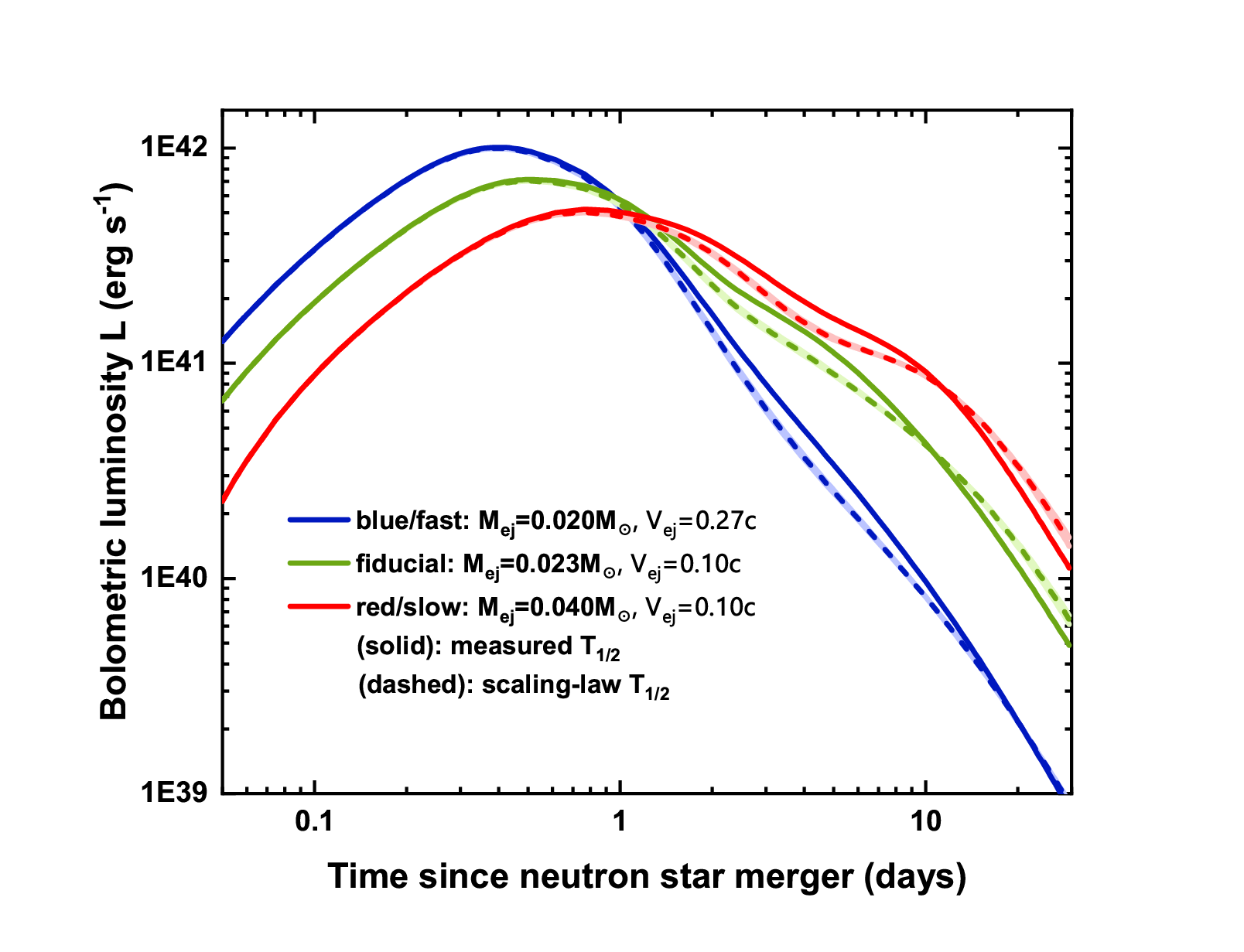}
\includegraphics[width=0.49\textwidth]{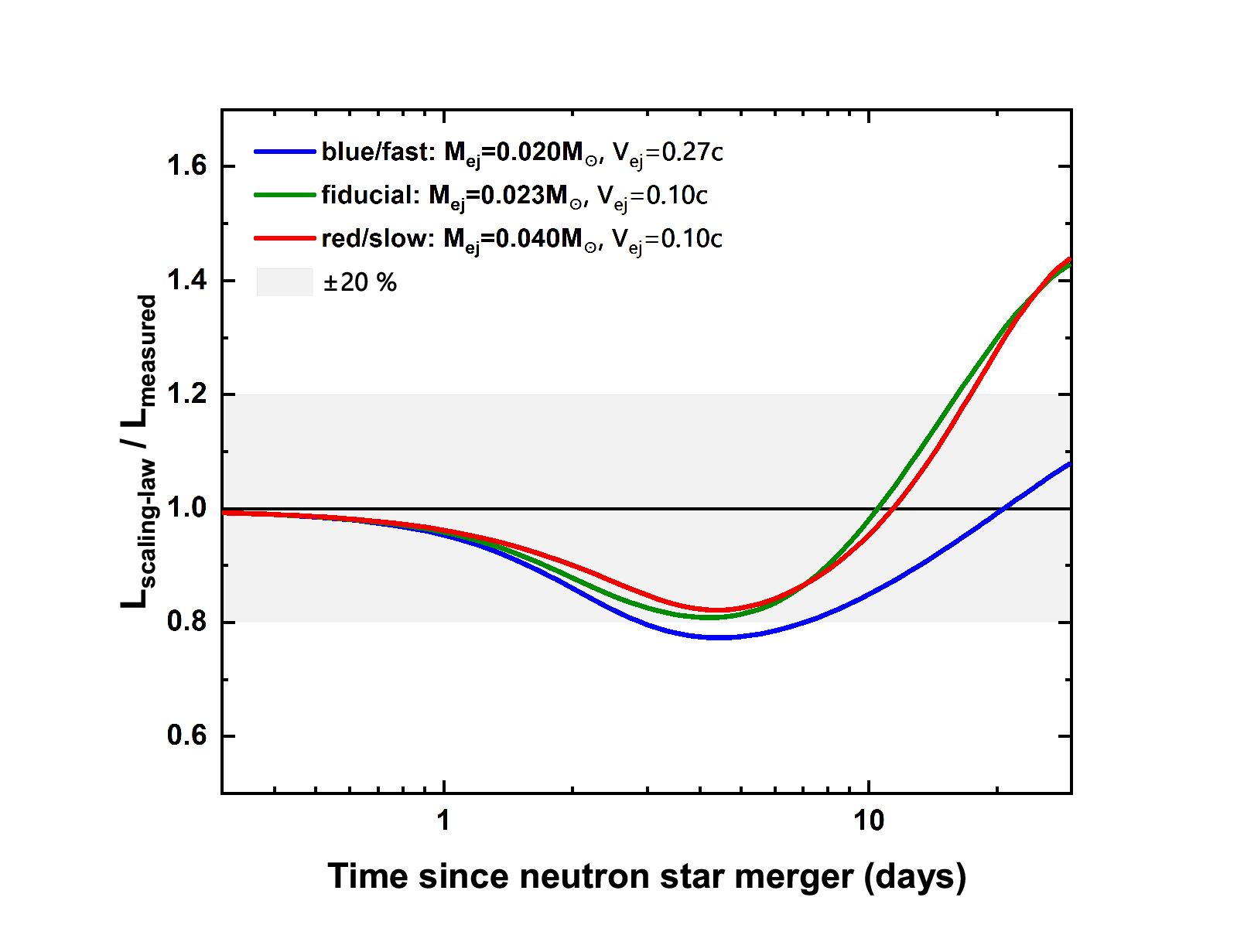}
\caption{(a) The simulated light curves for measured- (solid) vs.\ scaling-law-input (dashed) half-lives, repeated across three ejecta configurations spanning the blue/fast to red/slow range motivated by two-component fits to AT2017gfo~\cite{villar2017,cowperthwaite2017}. (b) The resulting model-agreement ratio in each configuration; the two half-life inputs remain broadly consistent throughout, with the specific level of agreement depending on the ejecta parameters.}
\label{fig:sensitivity}
\end{figure}

To evaluate the scaling law across the diverse hydrodynamic configurations of neutron-star mergers, we extended our calculations to two additional, GW170817-motivated cases bracketing the fiducial model: a fast, lanthanide-poor ``blue'' component ($M_{\rm ej}=0.020\,M_\odot$, $v_{\rm ej}=0.27c$, $\kappa \approx 0.5$--$1.0$ cm$^2$ g$^{-1}$) and a slow, lanthanide-rich ``red'' component ($M_{\rm ej}=0.040\,M_\odot$, $v_{\rm ej}=0.10c$, $\kappa \approx 1.0$--$10$ cm$^2$ g$^{-1}$).

Figure~\ref{fig:sensitivity} demonstrates that the simulated light curves remain in broad agreement across all three configurations. The blue/fast case agrees most tightly ($0.75$--$1.05\times$ over 1--29 days), while the fiducial and red/slow cases range somewhat further ($0.78$--$1.31\times$ and $0.80$--$1.32\times$, respectively). The scaling-law-input curve consistently runs dimmer at 2--5 days and brighter beyond $\sim 15$ days in every configuration. This confirms that the $Z^{2/3}/\sqrt{Q_\alpha}$ scaling variable remains robust across representative ejecta configurations for modeling the electromagnetic counterparts of gravitational-wave sources.

The robustness of this scaling law has direct implications for the inference of ejecta properties from gravitational-wave events. Standard kilonova models often approximate or neglect late-time $\alpha$-decay heating due to nuclear uncertainties, which can bias the posterior distributions of the ejecta mass ($M_{\rm ej}$) and opacity ($\kappa$) extracted from fitting AT2017gfo~\cite{villar2017,cowperthwaite2017}. By providing a physically grounded framework for the $\alpha$-decay component, the $Z^{2/3}/\sqrt{Q_\alpha}$ variable allows for the rigorous inclusion of late-time heating without introducing ad hoc nuclear parameters. This effectively decouples the macroscopic radiative-transfer modeling from the microscopic nuclear uncertainties, thereby reducing systematic errors in the multimessenger characterization of neutron-star merger ejecta.

\section{Conclusion}
We have established a universal $Z^{2/3}/\sqrt{Q_\alpha}$ scaling law for $\alpha$-decay half-lives across the measured even--even nuclear chart. The exponent $a\simeq2/3$ emerges from a uniquely sharp global optimization, reflecting the dimensional reduction of quantum tunneling and nuclear surface physics across the correlated heavy-nucleus manifold. Its persistence across independent microscopic and semi-empirical descriptions, without modification of their original parameters, provides strong cross-model support for $Z^{2/3}/\sqrt{Q_\alpha}$ as an emergent reduced coordinate for $\alpha$-decay systematics. Extending this scaling to an identified r-process $\alpha$-decay path yields a quasi-power-law radioactive heating rate consistent with published full-network results. Independent numerical calculations of the thermalized heating rates and radiative-diffusion light curves demonstrate that integrating our universal scaling predictions directly into the nuclear source terms yields multimessenger observables that accurately reproduce the kilonova AT2017gfo associated with the gravitational-wave event GW170817. This framework bridges fundamental quantum tunneling and multimessenger astrophysics, offering a robust tool for modeling the radioactive engines of neutron-star mergers, reducing systematic uncertainties in ejecta inference, and providing a foundation for extrapolating heavy-nucleus decay properties into experimentally inaccessible regions.

\end{document}